# Estimating resilience of annual crop production systems: theory and limitations.

*Zampieri M., Weissteiner C. J., Grizzetti B., Toreti A., van den Berg M. and F. Dentener*

*European Commission, Joint Research Centre (JRC), Ispra, Italy.*

*matteo.zampieri@ec.europa.eu*

*1) Agricultural production is sensitive to climate anomalies and extremes, which are increasing in frequency and intensity because of global warming, and generally to environmental and socio-economic adverse events. This motivates the need of a proper evaluation of the resilience of the agricultural production systems to enhance food security, market stability, and the capacity of society to cope with the effects of climate change.*

*2) Here, we investigate the problem of quantifying resilience in a simplified framework, focusing on the crop production component of the agricultural system. We propose a simple indicator of crop resilience to external shocks such as climate extremes and environmental and socio-economic stresses, simply computable on annual crop production time-series. The definition of this indicator is derived mathematically from the ecological definition of resilience applied to an idealized crop production system.*

*3) Through numerical experiments conducted with a conceptual crop model, we show the properties of the crop resilience indicator for production systems at different levels of adaptation to moderate climate anomalies, under different frequencies of extreme shocks, and accounting for simple socio-economical impacts of the production losses as well.*

*4) We discuss the applicability of the proposed approach to real agricultural production systems and the effects of crop diversity on the resilience of crop production systems.*

*5) Our findings suggest that the increase of frequency of extreme events affecting crop yields in the context of climate change is affecting proportionally the resilience of the cropping systems according to their levels of adaptation and management. In absolute terms, the crop production systems that are already at the optimal level of adaptation, maximizing the resilience for the current climatic conditions, are going to lose more resilience than the poorly adapted systems because of climate change. On the other hand, adaptation*

*can counterbalance the negative effects of climate change on the resilience of cropping systems with lower level of management.*

*6) Assessing the resilience of the crop production systems to adverse events and understanding the role of crop diversity using simple indicators as the one proposed in this study is relevant to plan strategies for facing climate change.*

# Introduction

The original meaning of the term resilience - firstly introduced in ecology – refers to the largest pressure that a system can cope with without changing its internal structure and losing its functioning capacity (Holling, 1973, 1996). Afterwards, the concept of resilience has spread also into other fields, such as engineering and social sciences, acquiring different definitions (Angeler & Allen, 2016; Berbés-Blázquez, Mitchell, Burch, & Wandel, 2017; Brand & Jax, 2007; Folke, 2006; Quinlan, Berbés-Blázquez, Haider, & Peterson, 2016). In the climate change context, a general definition of resilience is summarized by IPCC (IPCC, 2014) as "the capacity of social, economic, and environmental systems to cope with a hazardous event or trend or disturbance, responding or reorganizing in ways that maintain their essential function, identity, and structure, while also maintaining the capacity for adaptation, learning, and transformation".

Climate extremes, such as heat waves, droughts, exceptional rainfalls, frosts or hailstorms, and related unfavorable events, such as floods, fires, pest diseases or socio-economic crisis, are all hazardous events that can affect the agricultural crop production, causing partial or total failure of the annual yield. Crops are indeed able to resist and recover from many unfavorable weather events and other environmental stresses that characterize the local climate regimes. Crop production resilience can be enhanced by adaptation i.e. farm management practices that targets the most common stressors in each region by selecting resistant crop varieties, fine-tuning of crop calendars, proper management of soil, water, nutrients and pesticides, also through biodiversity-based solutions.

However, externally forced shocks of the crop production systems such as late spring frost (Barranco, Ruiz, & Gómez-del Campo, 2005; Eccel, Rea, Caffarra, & Crisci, 2009; Sweeney, Steigerwald, Davenport, & Eakin, 2013), heat stress and drought at flowering (Barnabàs, Jäger, & Fehér, 2007; Descamps, Quinet, Baijot, & Jacquemart, 2018), total failure of irrigation systems (Zampieri et al., 2018), spread of alien pests and pathogens (Lamichhane et al., 2015), heavy precipitation or flooding that impede crop and seed development or harvesting activities (Lesk, Rowhani, & Ramankutty, 2016; Shaw, Meyer, McNeill, & Tyerman, 2013) can result in quite large production losses.

Moreover, today's agriculture has been adapted to a range of conditions characterizing local climate regimes that are changing because of global warming (Iizumi & Ramankutty, 2016; Lobell, Schlenker, & Costa-Roberts, 2011; Ray, Gerber, Macdonald, & West, 2015; Zampieri, Ceglar, Dentener, & Toreti, 2017). Heat waves and droughts are becoming more common because of climate change (Gourdji, Sibley, & Lobell, 2013; Tebaldi & Lobell, 2018; Zampieri et al., 2016), as well as heavy precipitation events (Scoccimarro, Gualdi, Bellucci, Zampieri, & Navarra, 2013; Toreti et al., 2013). Therefore, in order to satisfy the increasing demand for food under the current trends of climate change (Foley et al., 2011), agricultural systems are likely to become less resilient if they don't adapt to changes in average conditions and to larger inter-annual fluctuations and increasing extreme conditions (Carr et al., 2016; Esper et al., 2017; Puma, Bose, Chon, & Cook, 2015).

This motivates the need for a proper characterization of resilience of agricultural production and food supply systems (Barrett, 2010; Suweis, Carr, Maritan, Rinaldo, & D'Odorico, 2015). Agricultural resilience and sustainability are often assessed together through holistic frameworks accounting for socio-economic, biophysical robustness and production diversity indicators (Seekell et al., 2017). Extensive literature reviews identified 30 relevant agro-ecosystem based sustainability indicators for climate resilient agriculture (Srinivasa Rao et al., 2018) and 15 different tools to assess resilience (Douxchamps, Debevec, Giordano, & Barron, 2017). In particular, FAO developed a complex framework accounting for a larger number of indicators (Russo & D'Errico, 2016). The EU proposed a more general conceptual framework defining the different aspects of resilience across agriculture and other sectors (Navracsics et al., 2015).

The aim of the present study is to contribute to the existing resilience assessment frameworks improving the understanding and the measurement of the component of resilience related to annual crop production. Previous resilience analyses involved food availability indicators based on the mean and the inter-annual variability of crop production (Gil, Cohn, Duncan, Newton, & Vermeulen, 2017; Kahiluoto et al., 2019; Khumairoh, Lantinga, Schulte, Suprayogo, & Groot, 2018; Srinivasa Rao et al., 2018; Zimmerer & De Haan, 2017).

Here, applying the original ecological definition of resilience (Holling, 1973) to crop production systems, we can define the crop resilience indicator ($R_c$). We discuss the general properties of $R_c$ using numerical simulations conducted with a conceptual crop model accounting for different levels of adaptation and management, for the effects of climate change, and for simple socio-economical consequences of the crop production losses as well. Finally, we discuss the limitations of the proposed indicator and its applicability to the production time-series recorded on larger and/or diversified spatial aggregations characterizing the real agricultural systems.

## Material and Methods

**Definition of the crop resilience indicator**

We consider an idealized agricultural production system composed of a single annual crop. This crop is grown on a generic spatial unit that is small enough to respond homogeneously to the pressures exerted by weather and climatic events, which constitute the external forcing of the production system. We also assume that the crop production response to the forcing is stationary (i.e. time-independent) and that the system bears no memory of the previous year conditions and that the effect on production have no repercussion on the following year i.e. the production time-series are independent and identically distributed.

Since the larger departures from the optimal climatic conditions for the crop growth are more rare, we measure the severity of the annual forcing through the return periods ($T^*$) of the climate anomalies, expressed in years. Using the return period allows accounting for shocks coming from different origin, such as different types of climate anomalies and other adverse events for the crop. It is not related to the concept of restoring time to the equilibrium conditions after a shock that is adopted by the engineering definition of resilience (Holling, 1996) (which cannot be applied on annual crop production data).

Following the ecological definition of resilience ($R_e$) as "the largest disturbance that a system can absorb before it loses its normal functioning" (Holling, 1973), we define the resilience of the crop production system as the largest departure from the optimal conditions that the crop production system can sustain without losing its capacity to produce the annual yield, measured by the return period ($T^*$):

(1) $R_e \equiv T^*_{MAX}$,

where $T^*_{MAX}$ is the return period of the maximum forcing that the crop can tolerate before losing completely the ability to produce the yield. In a stationary framework, $T^*_{MAX}$ is equivalent to the inverse probability of crop failure ($F$), leading to total production loss:

(2) $T^*_{MAX} = 1 / F$.

Now, we consider the special case of annual crop production time-series where the production values recorded at year $j$ ($p_j$) are either the potential ($p_j = P$) or zero ($p_j = 0$). This is equivalent to an idealized crop production system that is optimally adapted to the local climate regime (like, for instance, intensively managed systems with fully irrigated, fertilized and protected crops), optimized to be unaffected by moderate environmental stresses ($T^* < T^*_{MAX}$), but still sensitive to extreme events exceeding a certain threshold ($T^* > T^*_{MAX}$). We note that at large aggregated spatial scales (i.e. country or regional scales), total yield losses are very rarely observed, but they can occur more frequently at the field scale and for small spatial aggregations.

The mean and the variance of such time-series are, respectively:

(3) $\mu = P (1 - F)$, and

(4) $\sigma^2 = P^2 (1 - F) F$.

For this particular crop production system, it is possible to derive an indicator that is mathematically consistent with the ecological definition of resilience: combining equation and 3 and 4, equation 5 eliminates $P$ and related the annual production mean and variance to total production loss probability:

(5) $\mu^2 / \sigma^2 = (1 - F) / F$.

For F << 1, the following approximation holds: $(1 - F) / F \simeq 1 / F$.

Finally, using equation 1 and 2, we obtain the indicator for crop resilience ($R_c$):

(6) $R_c \equiv \mu^2 / \sigma^2 = R_e$,

$R_c$ increases with the mean annual production and decreases with its variance, which is in agreement with the common understanding of resilience and in line both with the farmer goals and with the expectations of the sectors of economy and society that come after the agricultural production in the food cycle.

**Design of numerical experiments**

We perform numerical experiments with a more realistic – albeit still idealized – crop model consisting of different crop production damage functions that characterize the production effects response to a randomly generated annual external forcing:

(7) $p_{ij} = P \cdot (1 - D_i(T^*_j, T^*_{MAX}))$,

where $D_i$ is the damage function, which depends on the return period of the external forcing occurring on year $j$ ($T^*_j$), the return period of the maximum forcing that the crop system can cope with without losing completely the yield ($T^*_{MAX}$), and the adaptation level ($i$).

This conceptual model allows computing the $R_C$ indicator for cropping systems varying the frequencies of extreme events causing total yield loss ($T^*_{MAX}$), the level of adaptation to moderate environmental stresses ($i=low,mid,high$), and including impacts lasting more than one cropping season as well to account for simple socio-economic effects (defined later by equation 8).

The external forcing statistical distribution is stationary, which allows producing homogeneous time-series long enough to compute $R_c$ reliably. The sampling error of $R_C$ is computed from Monte Carlo simulations of production time-series with different length $n$ and $\sigma / \mu$ ratios.

Figure 1 show the prescribed statistical probability distribution function (PDF, black thick line) of the external forcing annual time-series as a function of the return periods for six examples of damage functions: two values of $T^*_{MAX}$ and three levels of adaptation ($m=low,mid,high$ in red,green,blue, respectively).

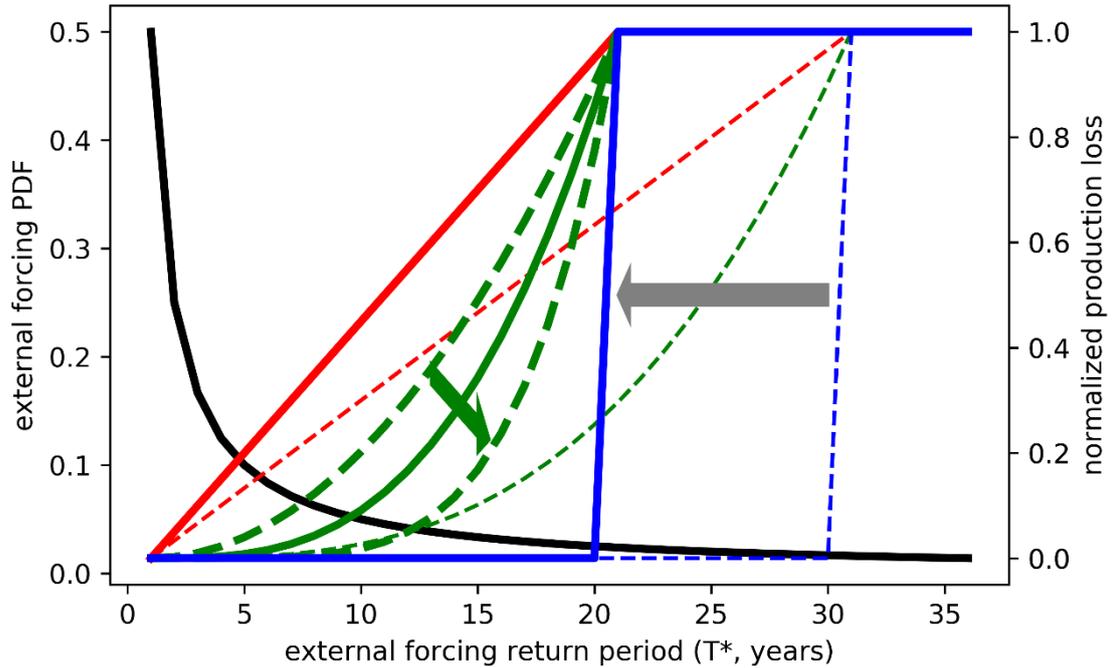

*Figure 1:* Probability Density Function (PDF) of the external forcing (black line, left axis) plotted as a function of the return periods ($T^*$). The PDF is defined by a gamma function (shape = 1, scale = 2). Annual crop production damage functions (right axis) for specified total production loss return periods ($T^*_{MAX}$ = 20 years, bold lines; $T^*_{MAX}$ = 30 years, thin dashed lines). The damage functions for return periods smaller than $T^*_{MAX}$ represent three different crop production responses to moderated external forcing: linear, polynomial (i.e. third power) and stepwise (Heaviside) in red, green and blue, respectively, representing low, moderate and high levels of adaptation and management of the cropping system (i=low,mid,high). Bold dashed green lines represent trajectories of damages following increasing levels of adaptation and management (green arrow), which minimizes the losses from moderated external forcing return periods. The grey arrow represents the effect of climate change that increases the frequency of extreme climate events leading to total crop production loss.

Small values of $T^*$ compared to $T^*_{MAX}$ correspond to the normal climate regime consisting of optimal or sub-optimal conditions for the crop, which is reflected in limited production losses. Higher $T^*$ characterize extreme events that are detrimental for the crop and associated with large or eventually total yield loss if $T^*$ exceeds $T^*_{MAX}$.

The total yield loss return period ($T^*_{MAX}$) is used as free parameter to mimic the effects of different climatic conditions (and climate change) through the crop damage functions. The damage functions in Figure 1 are representative for two selected $T^*_{MAX}$ values (20 and 30 years), chosen as illustrative examples. Total yield loss occurs for external forcing characterized by return period larger than these thresholds ($T^* > T^*_{MAX}$, normalized production loss equal to 1).

Below these thresholds (i.e. more frequently occurring climate events and environmental stresses), the loss can be different according to different types of production systems. The dependency on moderate environmental stresses is simulated by different shapes of the crop damage function for $T^* < T^*_{MAX}$. Here we consider three cases, represented by the coloured solid lines in Figure 1:

1) the linear damage function (red line), typically representing low level of adaptation and low management intensity (*i=low*) as it produces significant production losses also for very common environmental stresses and departures from the optimal climatic conditions ($T^* << T^*_{MAX}$);

2) the stepwise damage function (Heaviside function, blue line) that was used to derive $R_c$ (eq. 6), representing system with a high level of adaptation and high management intensity (*i=high*) that are sensitive only to extreme events happening only at return periods larger or equal $T^*_{MAX}$ and not to the more common environmental stresses;

3) an intermediate polynomial function (third power, green line), which may represent a more realistic production response to climatic variability (moderate adaptation level, *i=mid*).

As it is evident from Figure 1, changing values of $T^*_{MAX}$ affect the production losses for moderate external forcing as well.

In the next section we show the results of the crop resilience indicator computed on such production time-series for different levels of adaptation and different $T^*_{MAX}$. In addition, production time-series are simulated accounting for cropping systems where the impacts of the external forcing persist longer than one season, which could actually happen in real agricultural systems because of socioeconomic reasons as, for instance, reduced investments capacity. These cases are simulated using the same damage functions described before, but summing to the loss that the external forcing would produce in each year, half of the loss that the external forcing would produce in the previous year:

(8) $D'_{ij} = D_{ij} + D_{i,j-1}/2$

Different impacts and longer timescales could be considered, but are not shown.

## Results

**Estimation of Crop Resilience for different climates and adaptation levels**

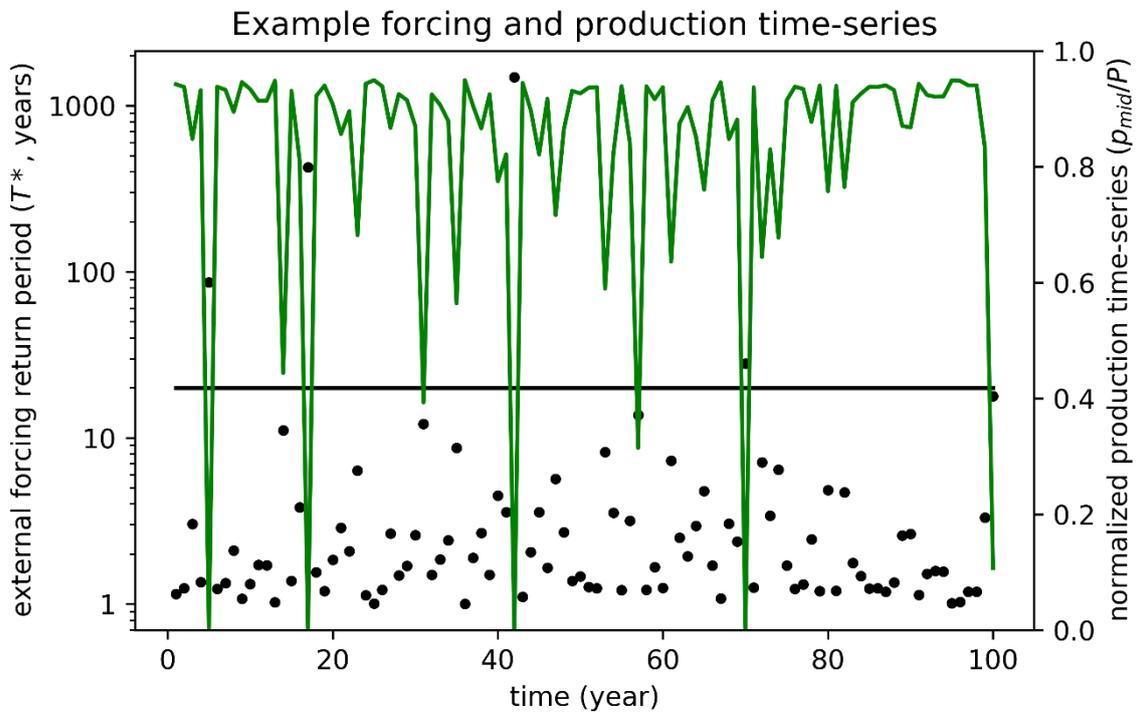

*Figure 2: Example time series of external forcing (dots, expressed as the corresponding return periods T\*) and production (green lines, normalized) obtained with a polynomial damage function (corresponding to moderate adaptation level, i=mid) and total yield loss return period of 20 years. The horizontal line represents the critical threshold of the forcing that produces total yield loss in this virtual agro-climatic realization.*

Figure 2 shows a randomly generated time-series of the external forcing (black dots) following the prescribed PDF (black line in Figure 1) and the corresponding annual crop production time-series resulting from the polynomial damage function (green line in Figure 1). In this example, the critical threshold of the return period is set to 20 years ($T^*_{MAX} = 20$, horizontal line), meaning that total production loss is expected to be recorded on average every 20 years (it occurs 4 times in the random realization shown in Figure 2). The crop resilience computed for the production time-series plotted in Figure 2 is equal to 11.2.

Figure 3 summarizes the same result computed for the different crop damage functions depicted in Figure 1, for a wide range of return periods of total production loss ($T^*_{MAX}$ = 10 to 100 years), and including impacts lasting longer than one season.

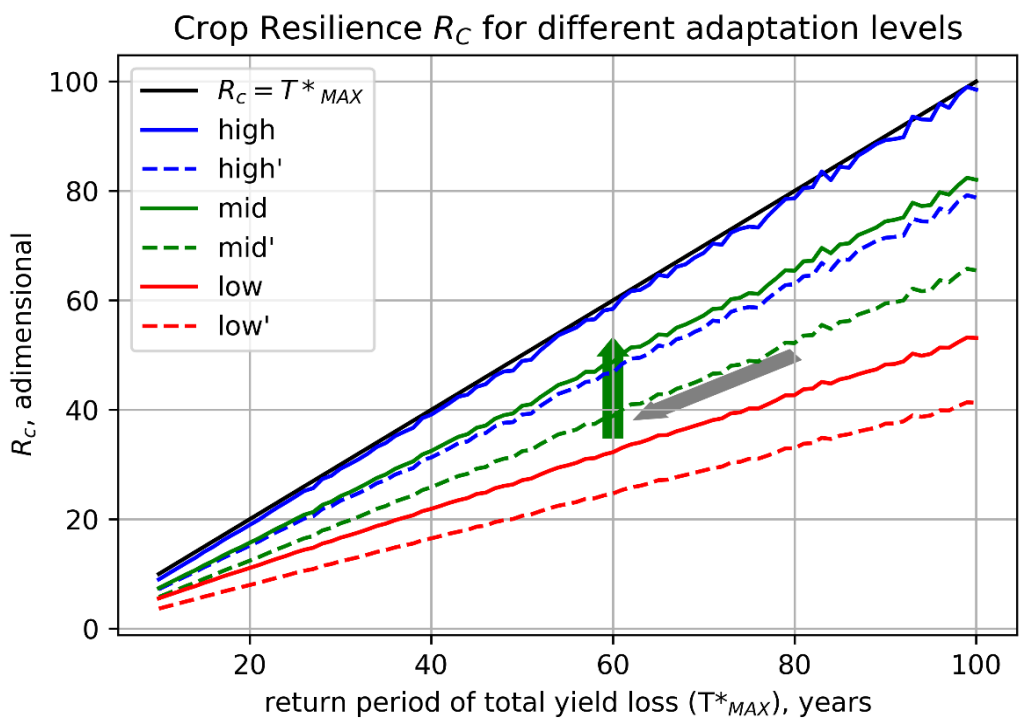

*Figure 3: Crop Resilience ($R_c$) computed for return levels ranging from 10 to 100 years and for different damage functions: Heaviside function representing high level of adaptation (i=high, blue lines), polynomial function for moderate adaptation (i=mid, green lines), linear function for poorly managed crops (i=low, red lines). Dashed lines show the estimated $R_c$ for cropping systems influenced by the external forcing occurred previous year (high', mid', low', see text for explanation). As in Figure 1, the arrows show the direction of the effects of climate change (grey) and adaption (green) on $R_c$.*

Figure 3 shows the $R_c$ computed on crop production time-series for cropping systems characterized by different return periods of total yield loss ($T^*_{MAX}$) and different levels of adaptation and management. In all cases, near-linear relationships between $R_c$ and return period of total yield loss ($T^*_{MAX}$) are found. The proportionality factor of the relationships decreases in crop production systems with lower adaptation level and bearing memory of the climate impact of the previous year. For instance, for a maximum yield loss return period of 40 years, the $R_c$ computed for the high adaptation system (*high*, blue line) is almost identical to $T^*_{MAX}$ (as expected). Adding to each year half of the damage produces in the previous year (*high'*) would lead to an $R_C$ of about 30. About the same $R_c$ is recorded for mid-level of adaptation and no memory (*mid*). In case of high sensitivity to moderate environmental stresses (i.e. low adaptation, *low*), $R_c$ is about half the value of the highly adapted and intensively managed agricultural system (*high*).

The effects of climate change and of increasing adaptation level can be thought as moving the production system along and across the lines represented in Figure 3 by the grey and green arrows, respectively. In absolute terms, resilience of crop production systems with low adaptation and poor management is comparatively less affected by changes in the frequency of extreme events with respect to the more developed systems because the production variability is largely determined by the impacts of smaller perturbations from the optimal climate conditions. Therefore, the impacts of climate change will be probably easier to detect in the more developed cropping systems. However, the relative effect is the same.

**Accuracy of crop resilience estimation**

Crop production resilience is difficult to estimate from limited duration time series. The relative error of $R_c$ depends on the number of observations and on the particular distribution of climate impacts i.e. on the $R_c$ itself. Table 1 shows the relative $R_c$ error ($\sigma_{Rc} / R_c$) computed from Monte Carlo simulations i.e. computing $R_c$ on a large number of random production time-series realizations with different lengths *n* and different σ / μ ratios, assuming Gaussian distribution of the crop production anomalies around the mean.

**Table 1:** *Relative uncertainty (in %) of $R_c$ computation on crop production time-series with different number of years (n) and different $\sigma/\mu$ ratios.*

| $\sigma_{Rc} / R_c$ | $\sigma/\mu$ = 5% ($R_c$ = 400) | $\sigma/\mu$ = 10% ($R_c$ = 100) | $\sigma/\mu$ = 20% ($R_c$ = 25) | $\sigma/\mu$ = 30% ($R_c$ =11.1) | $\sigma/\mu$ = 50% ($R_c$=4) |
|---|---|---|---|---|---|
| n= 10 | 63% | 64% | 65% | 67% | 73% |
| n= 20 | 37% | 38% | 38% | 39% | 44% |
| n= 30 | 28% | 29% | 29% | 30% | 34% |
| n= 40 | 24% | 24% | 25% | 26% | 29% |
| n= 50 | 21% | 21% | 22% | 23% | 26% |
| n= 100 | 15% | 15% | 15% | 16% | 18% |

The accuracy of $R_c$ estimation (Table 1) highly depends on the uncertainty related to the estimation of the standard deviation, which appears at the denominator in equation (6). Therefore, it is of primary importance to account for the uncertainties associated with $R_c$ computed from production time-series of limited duration, to derive meaningful information from real data. From Table 1, we also note that it is more difficult to estimate crop resilience for low resilience systems, but this is relatively less important compared to the dependency of the accuracy on the data availability, which is the more important limiting factor for accurate estimation of crop resilience.

**Crop diversity and resilience of crop production systems.**

In the previous sections, we have considered a homogeneous cropping system, exposed to stationary external forcing. In real-world conditions, however, crop production data is usually provided for larger inhomogeneous spatial units and/or for production systems characterized by different crops and crop varieties. We account for this issue by considering two types of situations, where crop responses to different climatic conditions are correlated, anti-correlated or uncorrelated.

The properties of $R_c$ computed for the sum of crop productions under such conditions can be understood by induction considering two crop (*k=1,2*) and using the following statistical relationships linking the sum of two variables:

(9a)     $p_{j,TOT} = p_{j,k=1} + p_{j,k=2}$

(9b)     $Ave(p_{TOT}) = Ave(p_{k=1}) + Ave(p_{k=2})$

(9c)     $Var(p_{TOT}) = Var(p_{k=1}) + Var(p_{k=2}) + 2 \cdot Covar(p_{k=1}, p_{k=2})$,

and then generalizing for an arbitrary number of crops. The production time-series that are integrated may be related to different crops or different crop varieties, but also to the same crop under different climatic regimes and/or adaptation levels.

For simplicity, we consider an aggregated spatial unit composed of two homogeneous areas with two crops that are statistically equivalent to each other i.e. they are characterized by the same production mean and variance. The $R_C$ indicator computed for the sum of these two production time-series is characterized by the following particular cases depending of the cross-correlation of the production time-series:

- if the crop production responses to a climate signal are perfectly correlated, the sum of the production is equal to twice one of the two; therefore, the mean and the standard deviation of the sum of the productions are doubled w.r.t. those of the individual production time-series; consequently, $R_c$ does not change w.r.t the $R_c$ computed on the individual production time-series ($R_{C,TOT} = R_{C,k}$).
- If the crop productions are perfectly anti-correlated, then the negative fluctuations of one spatial unit of crop production are fully balanced by positive fluctuations of the other crop spatial unit, so that the resulting standard deviation of the summed production is zero and $R_c$ is infinity: the crop production system has become perfectly resilient ($R_{C,TOT} = inf$).
- If the crop productions are not correlated, then mean and the variance of the sum are doubled w.r.t those of the individual production time-series (see equation 9c), so that $R_c$ of the total production will double as well ($R_{C,TOT} = 2 \cdot R_{C,k}$).

In an agricultural system composed of an arbitrary number of equivalent crops ($k = 1, 2 … N$), i.e. characterized by the same production mean and variance, but uncorrelated in time, the crop resilience indicator for the total production is equal to the crop resilience indicator of the individual crops multiplied by the number of crops ( $R_{C,TOT} = N \cdot R_{C,k}$). This can be defined as the *diversity theorem of the crop resilience indicator*, following as a natural consequence of the definition given in equation (6).

For unequal production levels, crops contribute proportionally to the overall crop system resilience. The same happens if the production time-series are partially cross-correlated.

The same general reasoning holds in case the aggregation of the time-series is performed on the nutritional annual production values (e.g. energy or protein content) or on the economic values of the annual production of the different crops.

## Discussion.

This paper addresses the issue of measuring crop production resilience from annual production time-series. The proposed framework considers the probability of occurrence of adverse events hitting crop production. For an idealized crop production system composed of a single crop, with no rotation, grown on a spatial unit responding homogeneously to external forcings, the crop resilience indicator ($R_c$), defined as the ratio between the squared mean and the variance of production time-series ($R_c \equiv \mu^2 / \sigma^2$), can be derived directly from the ecological definition of resilience (Holling, 1973).

The mathematical derivation of the crop resilience indicator is a main achievement of this study. Numerical simulations show that $R_c$ is in good approximation proportional to the return period of extreme events leading to total yield loss expressed in years ($T^*_{MAX}$). Climate change can be interpreted as the increase of frequency of such extreme events i.e. decreasing $T^*_{MAX}$, so the same proportional decrease is expected for crop production resilience.

We have found that the coefficient of proportionality between $R_c$ and $T^*_{MAX}$ is closely linked to the level of adaptation and management of the particular cropping system under evaluation (see Figure 3). In case of well adapted and highly managed production systems, which can cope with minor departures from the optimal conditions but are still sensitive to rarely occurring extreme events producing total yield loss, $R_c$ is exactly equal the total yield loss return period expressed in years ($R_c = T^*_{MAX}$).

$R_c$ decreases in less developed agricultural systems, because they are also influenced by more moderate deviations from the optimal climate and environmental conditions. $R_c$ is also lower in case of production impacts lasting longer than one season, which could happen in reality because of socioeconomic reasons (e.g. lack of financial resources to buy seeds for next year), for instance. For the less resilient crop production systems considered in our analysis, we found that the $R_c$ is about ½ · $T^*_{MAX}$. For real-world systems, the proportionality factor is probably between ½ and 1. In absolute terms, this means that climate change can affect more strongly the resilience of crop production systems under intensive management (because of the larger proportionality factor) with respect to the less-developed systems. We noted, however, that this ignores the fact that even small declines in the crop resilience of poorly or unmanaged crop production systems can result in harmful socio-economic effects on the farming system (Adger, 2006). According to the proposed framework, improving agro-management and adaptation increases the proportionality ratio between $R_c$ and $T^*_{MAX}$, potentially compensating the detrimental effects of climate change on crop production resilience of poorly developed cropping systems.

The computation of $R_c$ is relatively straightforward but also has limitations connected to the sample size (i.e. the length of the available production time-series to computed the index). For a resilient crop production system (i.e. $\sigma / \mu < 30\%$) with at least 30 years of production data, the uncertainty of the estimation ($\sigma_{Rc} / R_c$) is less than 30%. Larger sampling errors are associated to shorter time-series (see Table 1).

Several studies highlighted the importance of crop diversification to enhance production resilience (Davis, Hill, Chase, Johanns, & Liebman, 2012; Gil et al., 2017; Isaacs, Snapp, Chung, & Waldman, 2016; Martin & Magne, 2015; Meldrum et al., 2018; Prober & Smith, 2009). Consistently, the crop resilience indicator computed for spatially aggregated or diversified crop production systems characterized by uncorrelated or anti-correlated climate impacts is generally much larger than for homogeneous systems. In case of multiple crops with uncorrelated production time-series, total crop resilience scales proportionally to the number of equivalent crops (*diversity theorem of the crop resilience indicator*). This result strongly suggests that crop diversification, together with a consistent change of diet (Dwivedi et al., 2017), is indeed a viable strategy to limit the negative effects of increasing climate variability in the context of global warming.

When considering multiple cropping systems, it is also important to distinguish between the level of adaptation and the intensity of the agro-management. The first system can react to climate change by growing suitable and diversified crops and conserving biodiversity, the latter by increasing irrigation, fertilisation and pesticide inputs, which is clearly less sustainable (Zampieri et al., 2018).

This study provides the theoretical framework for several follow-up analyses addressing quantitative aspects of resilience and sustainability of real crop production systems using models and observed data for different world regions and at the global scale in the context of climate change.

**References**

Adger, W. N. (2006). Vulnerability. *Global Environmental Change*, *16*(3), 268–281. doi:10.1016/j.gloenvcha.2006.02.006

Angeler, D. G., & Allen, C. R. (2016). EDITORIAL: Quantifying resilience. *Journal of Applied Ecology*, *53*(3), 617–624. doi:10.1111/1365-2664.12649

Barnabàs, B., Jäger, K., & Fehér, A. (2007). The effect of drought and heat stress on reproductive processes in cereals. *Plant, Cell & Environment*, *31*(1), 11–38. doi:10.1111/j.1365-3040.2007.01727.x

Barranco, D., Ruiz, N., & Gómez-del Campo, M. (2005). Frost tolerance of eight olive cultivars. *HortScience*, *40*(3), 558–560.

Barrett, C. B. (2010). Measuring food insecurity. *Science (New York, N.Y.)*, *327*(5967), 825–828. doi:10.1126/science.1182768

Berbés-Blázquez, M., Mitchell, C. L., Burch, S. L., & Wandel, J. (2017). Understanding climate change and resilience: assessing strengths and opportunities for adaptation in the Global South. *Climatic Change*, *141*(2), 227–241. doi:10.1007/s10584-017-1897-0

Brand, F. S., & Jax, K. (2007). Focusing the meaning(s) of resilience: Resilience as a descriptive concept and a boundary object. *Ecology and Society*, *12*(1). doi:10.5751/ES-02029-120123

Carr, J. A., Odorico, P. D., Laio, F., Ridolfi, L., Odorico, D., Gephart, J. A., … Ratajczak, Z. (2016). Reserves and trade jointly determine exposure to food supply shocks Reserves and trade jointly determine exposure to food supply shocks.

Davis, A. S., Hill, J. D., Chase, C. A., Johanns, A. M., & Liebman, M. (2012). Increasing Cropping System Diversity Balances Productivity, Profitability and Environmental Health. *PLoS ONE*, *7*(10), 1–8. doi:10.1371/journal.pone.0047149

Descamps, C., Quinet, M., Baijot, A., & Jacquemart, A.-L. (2018). Temperature and water stress affect plant-

pollinator interactions in Borago officinalis (Boraginaceae). *Ecology and Evolution*, *8*(6), 3443–3456. doi:10.1002/ece3.3914

Douxchamps, S., Debevec, L., Giordano, M., & Barron, J. (2017). Monitoring and evaluation of climate resilience for agricultural development – A review of currently available tools. *World Development Perspectives*, *5*, 10–23. doi:10.1016/J.WDP.2017.02.001

Dwivedi, S. L., Bueren, E. T. L. Van, Ceccarelli, S., Grando, S., Upadhyaya, H. D., & Ortiz, R. (2017). Diversifying Food Systems in the Pursuit of Sustainable Food Production and Healthy Diets. *Trends in Plant Science*, *22*(10), 842–856. doi:10.1016/j.tplants.2017.06.011

Eccel, E., Rea, R., Caffarra, A., & Crisci, A. (2009). Risk of spring frost to apple production under future climate scenarios: the role of phenological acclimation. *International Journal of Biometeorology*, *53*(3), 273–286. doi:10.1007/s00484-009-0213-8

Esper, J., Büntgen, U., Denzer, S., Krusic, P. J., Luterbacher, J., Schäfer, R., … Werner, J. (2017). Environmental drivers of historical grain price variations in Europe. *Climate Research*, *72*(1), 39–52. doi:10.3354/cr01449

Foley, J. A., Ramankutty, N., Brauman, K. A., Cassidy, E. S., Gerber, J. S., Johnston, M., … Zaks, D. P. M. (2011). Solutions for a cultivated planet. *Nature*, *478*, 337. Retrieved from http://dx.doi.org/10.1038/nature10452

Folke, C. (2006). Resilience: The emergence of a perspective for social-ecological systems analyses. *Global Environmental Change*, *16*(3), 253–267. doi:10.1016/j.gloenvcha.2006.04.002

Gil, J. D. B., Cohn, A. S., Duncan, J., Newton, P., & Vermeulen, S. (2017). The resilience of integrated agricultural systems to climate change. *Wiley Interdisciplinary Reviews: Climate Change*, *8*(4), e461. doi:10.1002/wcc.461

Gourdji, S. M., Sibley, A. M., & Lobell, D. B. (2013). Global crop exposure to critical high temperatures in the reproductive period: Historical trends and future projections. *Environmental Research Letters*, *8*(2), 0–

10. doi:10.1088/1748-9326/8/2/024041

Holling, C. S. (1973). Resilience and stability of ecological systems. *Annual Review of Ecology and Systematics*, *4*, 1–23.

Holling, C. S. (1996). *Engineering Resilience versus Ecological Resilience*. The National Academy of Sciences.

Iizumi, T., & Ramankutty, N. (2016). Changes in yield variability of major crops for 1981 – 2010 explained by climate change. *Environmental Research Letters*, *11*(3), 0. doi:10.1088/1748-9326/11/3/034003

IPCC. (2014). Summary for Policymakers. In C. B. Field, V. R. Barros, D. J. Dokken, K. J. Mach, M. D. Mastrandrea, T. E. Bilir, … L. L. White (Eds.), *Climate Change 2014: Impacts, Adaptation, and Vulnerability. Part A: Global and Sectoral Aspects. Contribution of Working Group II to the Fifth Assessment Report of the Intergovernmental Panel on Climate Change* (pp. 1–32). Cambridge, United Kingdom, and New York, NY, USA: Cambridge University Press.

Isaacs, K. B., Snapp, S. S., Chung, K., & Waldman, K. B. (2016). Assessing the value of diverse cropping systems under a new agricultural policy environment in Rwanda. *Food Security*, *8*(3), 491–506. doi:10.1007/s12571-016-0582-x

Kahiluoto, H., Kaseva, J., Balek, J., Olesen, J. E., Ruiz-Ramos, M., Gobin, A., … Trnka, M. (2019). Decline in climate resilience of European wheat. *Proceedings of the National Academy of Sciences*, *116*(1), 123–128. doi:10.1073/pnas.1804387115

Khumairoh, U., Lantinga, E. A., Schulte, R. P. O., Suprayogo, D., & Groot, J. C. J. (2018). Complex rice systems to improve rice yield and yield stability in the face of variable weather conditions. *Scientific Reports*, *8*(1), 14746. doi:10.1038/s41598-018-32915-z

Lamichhane, J. R., Barzman, M., Booij, K., Boonekamp, P., Desneux, N., Huber, L., … Messéan, A. (2015). Robust cropping systems to tackle pests under climate change. A review. *Agronomy for Sustainable Development*, *35*(2), 443–459. doi:10.1007/s13593-014-0275-9

Lesk, C., Rowhani, P., & Ramankutty, N. (2016). Influence of extreme weather disasters on global crop

production. *Nature*, *529*(7584), 84–87. doi:10.1038/nature16467

Lobell, D. B., Schlenker, W., & Costa-Roberts, J. (2011). Climate Trends and Global Crop Production Since 1980. *Science*, *333*(6042), 616 LP-620. Retrieved from http://science.sciencemag.org/content/333/6042/616.abstract

Martin, G., & Magne, M. A. (2015). Agricultural diversity to increase adaptive capacity and reduce vulnerability of livestock systems against weather variability - A farm-scale simulation study. *Agriculture, Ecosystems and Environment*, *199*, 301–311. doi:10.1016/j.agee.2014.10.006

Meldrum, G., Mijatović, D., Rojas, W., Flores, J., Pinto, M., Mamani, G., … Padulosi, S. (2018). Climate change and crop diversity: farmers' perceptions and adaptation on the Bolivian Altiplano. *Environment, Development and Sustainability*, *20*(2), 703–730. doi:10.1007/s10668-016-9906-4

Navracsics, T., Sucha, V., Wahlstroem, M., Stigson, B., Wijkman, A., Lechner, S., … Jacometti, J. (2015). *The challenge of resilience in a globalised world*. JRC Science for Policy Report. doi:/10.2788/771635

Prober, S. M., & Smith, F. P. (2009). Enhancing biodiversity persistence in intensively used agricultural landscapes: A synthesis of 30 years of research in the Western Australian wheatbelt. *Agriculture, Ecosystems and Environment*, *132*(3–4), 173–191. doi:10.1016/j.agee.2009.04.005

Puma, M. J., Bose, S., Chon, S. Y., & Cook, B. I. (2015). Assessing the evolving fragility of the global food system. *Environmental Research Letters*, *10*(2). doi:10.1088/1748-9326/10/2/024007

Quinlan, A. E., Berbés-Blázquez, M., Haider, L. J., & Peterson, G. D. (2016). Measuring and assessing resilience: broadening understanding through multiple disciplinary perspectives. *Journal of Applied Ecology*, *53*(3), 677–687. doi:10.1111/1365-2664.12550

Ray, D. K., Gerber, J. S., Macdonald, G. K., & West, P. C. (2015). Climate variation explains a third of global crop yield variability. *Nature Communications*, *6*, 1–9. doi:10.1038/ncomms6989

Russo, L., & D'Errico, M. (2016). *RIMA-II Resilience Index Measurement and Analysis - II*. Retrieved from http://www.fao.org/3/a-i5665e.pdf%0Ahttp://www.preventionweb.net/publications/view/48984

Scoccimarro, E., Gualdi, S., Bellucci, A., Zampieri, M., & Navarra, A. (2013). Heavy precipitation events in a warmer climate: Results from CMIP5 models. *Journal of Climate*, *26*(20). doi:10.1175/JCLI-D-12-00850.1

Seekell, D., Carr, J., Dell'Angelo, J., D'Odorico, P., Fader, M., Gephart, J., … Tavoni, A. (2017). Resilience in the global food system. *Environmental Research Letters*, *12*(2). doi:10.1088/1748-9326/aa5730

Shaw, R. E., Meyer, W. S., McNeill, A., & Tyerman, S. D. (2013). Waterlogging in Australian agricultural landscapes: a review of plant responses and crop models. *Crop and Pasture Science*, *64*(6), 549–562. Retrieved from https://doi.org/10.1071/CP13080

Srinivasa Rao, C., Kareemulla, K., Krishnan, P., Murthy, G. R. K., Ramesh, P., Ananthan, P. S., & Joshi, P. K. (2018). Agro-ecosystem based sustainability indicators for climate resilient agriculture in India: A conceptual framework. *Ecological Indicators*, (June), 0–1. doi:10.1016/j.ecolind.2018.06.038

Suweis, S., Carr, J. A., Maritan, A., Rinaldo, A., & D'Odorico, P. (2015). Resilience and reactivity of global food security. *Proceedings of the National Academy of Sciences*, *112*(22), 6902–6907. doi:10.1073/pnas.1507366112

Sweeney, S., Steigerwald, D. G., Davenport, F., & Eakin, H. (2013). Mexican maize production: Evolving organizational and spatial structures since 1980. *Applied Geography*, *39*, 78–92. doi:10.1016/j.apgeog.2012.12.005

Tebaldi, C., & Lobell, D. (2018). Differences , or lack thereof , in wheat and maize yields under three low-warming scenarios OPEN ACCESS Differences , or lack thereof , in wheat and maize yields under three low-warming scenarios.

Toreti, A., Naveau, P., Zampieri, M., Schindler, A., Scoccimarro, E., Xoplaki, E., … Luterbacher, J. (2013). Projections of global changes in precipitation extremes from Coupled Model Intercomparison Project Phase 5 models. *Geophysical Research Letters*, *40*(18). doi:10.1002/grl.50940

Zampieri, M., Ceglar, A., Dentener, F., & Toreti, A. (2017). Wheat yield loss attributable to heat waves,

drought and water excess at the global, national and subnational scales. *Environmental Research Letters*, *12*(6). doi:10.1088/1748-9326/aa723b

Zampieri, M., Garcia, G. C., Dentener, F., Gumma, M. K., Salamon, P., Seguini, L., & Toreti, A. (2018). Surface freshwater limitation explains worst rice production anomaly in India in 2002. *Remote Sensing*, *10*(2). doi:10.3390/rs10020244

Zampieri, M., Russo, S., di Sabatino, S., Michetti, M., Scoccimarro, E., & Gualdi, S. (2016). Global assessment of heat wave magnitudes from 1901 to 2010 and implications for the river discharge of the Alps. *Science of the Total Environment*, *571*. doi:10.1016/j.scitotenv.2016.07.008

Zimmerer, K. S., & De Haan, S. (2017). Agrobiodiversity and a sustainable food future. *Nature Plants*, *3*(4). doi:10.1038/nplants.2017.47